\documentclass{aastex631}
\usepackage{amsmath}

\shorttitle{Effects of the Galactic magnetic field on the UHECR correlation studies with starburst galaxies}
\shortauthors{Higuchi et al.}
\graphicspath{{./}{figures/}}

\begin{document}

\title{Effects of the Galactic magnetic field on the UHECR correlation studies with starburst galaxies}

\author{Ryo Higuchi}
\affiliation{Astrophysical Big Bang Laboratory, RIKEN, 2-1 Hirosawa, Wako, Saitama 351-0198, Japan}
\author{Takashi Sako}
\affiliation{Institute for Cosmic Ray Research, The University of Tokyo, 5-1-5 Kashiwanoha, Kashiwa, Chiba 277-8582, Japan}
\author{Toshihiro Fujii}
\affiliation{Osaka Metropolitan University, 3-3-138 Sugimoto Sumiyoshi-ku, Osaka-shi, 558-8585, Japan}
\author{Kazumasa Kawata}
\affiliation{Institute for Cosmic Ray Research, The University of Tokyo, 5-1-5 Kashiwanoha, Kashiwa, Chiba 277-8582, Japan}
\author{Eiji Kido}
\affiliation{Astrophysical Big Bang Laboratory, RIKEN, 2-1 Hirosawa, Wako, Saitama 351-0198, Japan}

\correspondingauthor{Ryo Higuchi}
\email{ryo.higuchi@riken.jp}

\begin{abstract}
We estimate the biases caused by the coherent deflection of cosmic rays due to the Galactic magnetic field (GMF) in maximum-likelihood analysis for searches of ultrahigh-energy cosmic ray (UHECR) sources in the literature.
We simulate mock event datasets with a set of assumptions for the starburst galaxy (SBG) source model \citep{Aab2018}, coherent deflection by a GMF model \citep{Jansson2012a, Jansson2012b}, and mixed-mass composition \citep{Heinze2019}; we then conduct a maximum-likelihood analysis without accounting for the GMF in the same manner as previous studies. 
We find that the anisotropic fraction $f_{\rm ani}$ is estimated systematically lower than the true value. 
We estimate the true parameters which are compatible with the best-fit parameters reported in \cite{Aab2018}, and find that except for a narrow region with a large anisotropic fraction and small separation angular scale a wide parameter space is still compatible with the experimental results. 
We also develop a maximum-likelihood method that takes into account the GMF model and confirm in the MC simulations that we can estimate the true parameters within a 1$\sigma$ contour under the ideal condition that we know the event-by-event mass and the GMF.
\end{abstract}

\section{Introduction} \label{sec:intro}
Cosmic rays (CRs) are high-energy nuclei that come throughout the universe. 
Specifically, ultrahigh-energy cosmic rays (UHECRs) with energies around $100\,\rm EeV$ ($10^{20}\,\rm eV$) are observed but their origin is not known yet. 
There are two leading experiments that observe UHECRs: the Telescope Array (TA) experiment \citep[located in the U.S.A., 39.3$\,\deg$ N, 112.9$\,\deg$ W]{Kawai2008, Sagawa2020} and the Auger experiment \citep[located in Argentina, 35.2$\,\deg$ S, 69.5$\,\deg$ W]{Aab2015c}
covering the sky in the northern and southern hemispheres, respectively. 
Thanks to a large number of UHECR events observed through these experiments in the last decade, the arrival directions of UHECRs are found to be anisotropic in the intermediate-angular (i.e.$\sim 10$ to $\sim 20 \deg$) scale \citep{Abreu2012, Aab2015b, Abbasi2014},
which is believed to give us keys to knowing the UHECR origins. 
Due to the energy loss through the photo-pion production or photo-nuclear dissociation, most UHECRs cannot propagate more than 30 -- 100 Mpc \citep[the GZK limit]{Greisen1966, Zatsepin1966}. 
Consequently as astronomical candidates of UHECR sources, it is natural to consider nearby extragalactic high-energy objects. 
Previous studies have investigated the correlation between the arrival directions of UHECRs and their source candidates \citep{Abreu2007, Abreu2010, He2016, Aab2018, Abbasi2018}. 
Especially, recent studies \citep{Abreu2010,Aab2018,Abbasi2018} try to explain the arrival direction of UHECRs by a weighted sum of events originating from sources and isotropic backgrounds (see also the review in \cite{Batista2019}).
In these studies, the flux of UHECRs (CR flux model) is composed of the source-originated flux (source flux model) and the isotropic backgrounds (isotropic flux model). 
Based on a catalog of source candidates, the source flux model is constructed as a superposition of the Gaussian-smeared angular distributions of an individual point source. 
Previous studies introduced the anisotropic fraction $f_{\rm ani}$ as the fraction of source contribution to the total CR flux and the separation angular scale $\theta$ as the scale of the Gaussian-smearing. 
The separation angular scale $\theta$ is considered to reflect the deflections and scattering by the Galactic and extragalactic magnetic fields (EGMF).
The parameters $(f_{\rm ani},\theta)$ are searched to best fit the observed CR angular distribution (see Section \ref{sec:method} for details).

One of the possible source flux models suggested by the previous studies is that UHECRs are originated from nearby starburst galaxies (SBGs).
\cite{Aab2018} investigated the correlations between UHECRs observed by the Auger experiment and the CR flux models constructed with nearby extragalactic high-energy objects such as SBGs, active-galactic nuclei, and gamma-ray bursts. 
They reported the best correlation was found above $39\,\rm EeV$ with the SBG source flux model (SBG model), and the best-fit anisotropic fraction $f_{\rm ani}^{\rm Auger}$ and the separation angular scale $\theta^{\rm Auger}$ were estimated to be $9.7 \,\%$ and $12.9 \,\deg$, respectively (we call this best-fit model as the Auger best-fit model).

The TA experiment studied the correlation between the UHECR arrival directions observed in the northern sky and the source flux model with SBGs using the best-fit parameters reported by \cite{Aab2018}. It concluded that the UHECR arrival directions are compatible with both the isotropic distribution ($f_{\rm ani}=0\%$) and the Auger best-fit model \citep{Aab2018} with current statistics \citep{Abbasi2018}. 
While the nearby SBGs are one of the attractive candidates for the UHECR sources, it is surprising that even this best-fit source model can explain only $\sim10\,\%$ of the observed UHECRs.
This question motivated us to study possible biases to underestimate $f_{\rm ani}$ and to estimate a realistic constraint of $f_{\rm ani}$ deduced from the observations.

The isotropic-scattering approximation in the previous studies does not reflect the actual structure of the GMF, which deflects UHECR trajectories in a certain direction (coherent deflection). 
Current analyses treating the coherent deflection as a part of isotropic scattering may result in a smaller $f_{\rm ani}$ and a larger $\theta$ than the true values. 
In this study, we call these systematic effects on the parameter estimation caused by the GMF ``the GMF bias". 
To consider the GMF bias (besides introducing the GMF model), we take into account the following two components.
1) Dependence on the arrival direction. 
The coherent deflection around the Galactic center (GC) and the Galactic plane (GP) is much larger than in other regions of the sky. 
This dependence also affects analyses whose samples are divided into the northern sky (TA experiment) and the southern sky (Auger experiment). 
Generally, UHECRs observed in the southern sky should be affected by the GMF more than those in the northern sky.
These effects caused by the limitation of the sky coverage of each experiment are not evaluated in the previous studies.
2) Dependence on the rigidity $R=pc/Ze \sim E/Ze$. 
Here $pc$, $E$, and $Ze$ represent the particle momentum, energy, and electric charge, respectively. The approximation is valid in the ultra-relativistic regime considered here. 
Because the deflection angle is proportional to $1/R$ of each CR, the magnetic field effect is strongly coupled with the energy spectrum and the mass composition.
The Auger experiment suggests that the mass composition of UHECRs becomes heavier at higher energy \citep{Aab2017a, Batista2019, Heinze2019}.

In this study, we investigate the GMF bias in previous studies applying a commonly-used GMF model \citep{Jansson2012a, Jansson2012b} by means of Monte Carlo (MC) simulations. 
We generate mock event datasets assuming the true parameters $(f_{\rm ani}^{\rm true},\theta^{\rm true})$ taking into account coherent deflections by the GMF and a mass model (a mixed-mass spectrum model proposed in \cite{Heinze2019}).
For some samples of mock events, we demonstrate that the event arrival directions are apparently displaced from the real source directions.
The size of the displacements strongly depends on the direction of the sky and the rigidity.
Then, we applied a maximum-likelihood analysis to the mock events in the same manner as the previous studies. 
In order to focus on the bias in the previous analysis \citep{Aab2018, Abbasi2018}, we fix the source flux model to be the SBG model.
Based on the analyses of the mock datasets, we discuss the biases in the parameter estimation separately in the northern, southern, and all-sky regions.
We also develop an analysis technique to reduce the GMF bias. 

\section{Mock Dataset Production} \label{sec:sample}
\begin{figure}
\epsscale{0.7}
\plotone{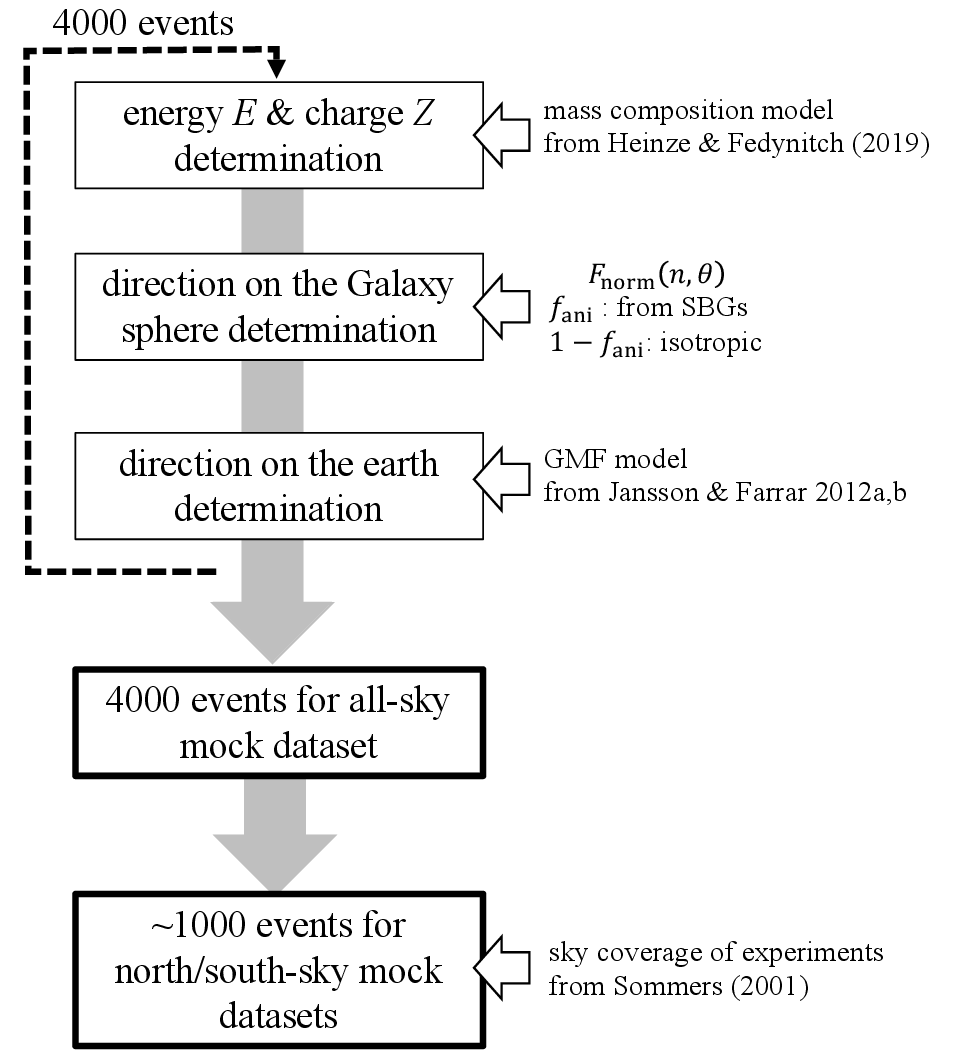}
\caption{
A schematic for a mock dataset generation.}
\label{fig:schematic}
\end{figure}

The mock event datasets are generated under a set of assumptions (the SBG model, the GMF model, the mass-dependent energy spectrum) with a flow shown in Figure \ref{fig:schematic}. 
The construction of the CR flux models before and after considering the GMF deflection is described in Section \ref{sec:fluxmodel} and \ref{sec:mod}, respectively.
To generate mock events taking into account the coherent deflections caused by the GMF, one needs to assign a rigidity $R$, {\it i.e.} energy and mass, to each event. 
We use a mixed-mass assumption to reflect a realistic situation as described in Section \ref{sec:genmmc}. 
Results with single-mass assumptions are summarized in Appendix \ref{sec:mockpure}.
In all cases, we generate 1000 datasets, each of which contains 4000 mock events across the whole sky (all-sky dataset). 
To compare the datasets with the observed UHECRs by the TA and Auger experiments, in Section \ref{sec:exp} we select the north-sky and south-sky datasets from the all-sky datasets taking into account the sky coverage of each experiment.

\subsection{CR flux model originated from the SBG model} \label{sec:fluxmodel}
\begin{figure}
\epsscale{0.7}
\plotone{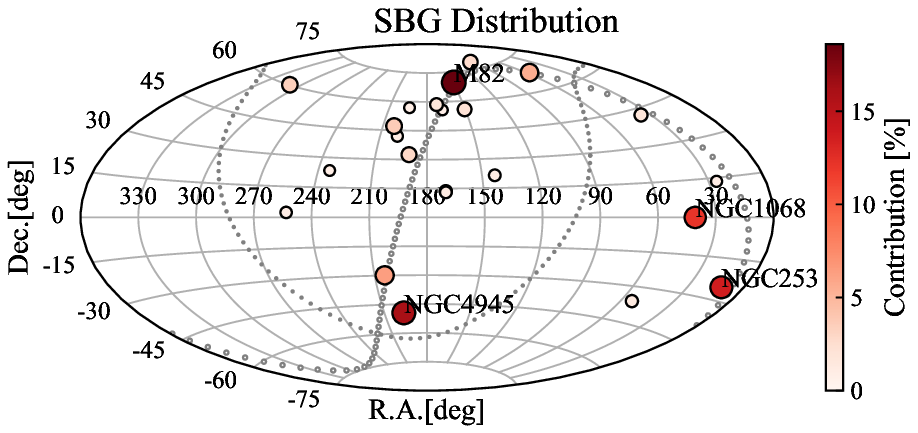}
\caption{
Directions and contributions of SBGs in Table \ref{table:aab2018tab1} from \cite{Aab2018} (in equatorial coordinates).
Circles show the direction of SBGs.
The color and area of each marker scale their relative flux contribution to the CR flux models.
The grey dots (circles) indicate the Galactic plane (the supergalactic plane).
\label{fig:sbg}}
\end{figure}

\begin{figure}
\epsscale{0.7}
\plotone{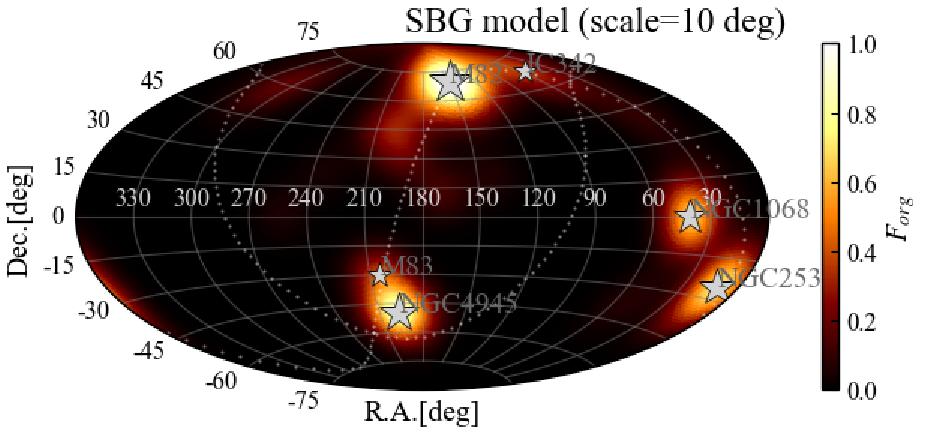}
\caption{
An example of the SBG model with $\theta =10 \,\deg$ (in equatorial coordinates).
The white dotted lines represent the Galactic plane (GP) and the supergalactic plane (SGP). 
The top six contributing SBGs are noted as grey stars. 
}\label{fig:f_g}
\end{figure}

With a source model fixed to the SBG model, the CR flux model is constructed with the assumption of a two-parameter set $(f_{\rm ani},\theta)$. 
In \cite{Aab2018} and \cite{Abbasi2018} the CR flux model $ F_{\rm org}({\bf n}, \theta)$ is determined as the superposition of the von Mises-Fisher function \citep[the Gaussian distribution on the sphere]{Fisher1953} of each source: 
 \begin{eqnarray}
 \label{eq:fmod2}
F_{\rm org}({\bf n},\theta)=\frac{\sum_{i} f_i \exp({\bf n}_i\cdot {\bf n} / \theta^2)}{{\int_{4\pi}{\sum_{i} f_i \exp({\bf n}_i\cdot {\bf n} / \theta^2)}d\Omega}}.
\end{eqnarray} 
Note that $i$ indicates each SBG, and $\bf n_i$ and $f_i$ mean its direction and relative flux (contribution from each source), respectively. 
Table \ref{table:aab2018tab1} is the list of SBGs in our SBG model defined in \cite{Aab2018}, which contains the values of $f_i$ and ${\bf n}_i$. 
The relative flux of SBGs $f_i$ in Table \ref{table:aab2018tab1} is determined by their continuum radio flux.
The source directions and relative contributions in Table \ref{table:aab2018tab1} are visualized in Figure \ref{fig:sbg}. 
We can see that most SBGs are located along the supergalactic plane (SGP)  and the top-4 contributions of SBGs dominate with $\sim 60 \,\%$ of the total flux.
\begin{deluxetable}{lcccc}
\tablecaption{Catalogue of SBGs in \cite{Aab2018}
\label{table:aab2018tab1}}
\tablewidth{0pt}
\tablehead{\colhead{ID$^1$} & \colhead{l [$^\circ$]$^2$} & \colhead{b [$^\circ$]$^2$} & \colhead{D [Mpc]$^3$} & \colhead{$f$ [$\%$]$^4$} }
\startdata
NGC 253  	&	97.4	&	-88	&	2.7	&	13.6 \\ 
M82  	&	141.4	&	40.6	&	3.6	&	18.6\\	
NGC 4945  	&	305.3	&	13.3	&	4	&	16\\
M83  	&	314.6	&	32	&	4	&	6.3\\
IC 342  	&	138.2	&	10.6	&	4	&	5.5\\
NGC 6946  	&	95.7	&	11.7	&	5.9	&	3.4\\	
NGC 2903  	&	208.7	&	44.5	&	6.6	&	1.1\\	
NGC 5055  	&	106	&	74.3	&	7.8	&	0.9\\	
NGC 3628  	&	240.9	&	64.8	&	8.1	&	1.3\\	
NGC 3627  	&	242	&	64.4	&	8.1	&	1.1\\	
NGC 4631  	&	142.8	&	84.2	&	8.7	&	2.9\\	
M51  	&	104.9	&	68.6	&	10.3	&	3.6\\	
NGC 891  	&	140.4	&	-17.4	&	11	&	1.7	\\
NGC 3556  	&	148.3	&	56.3	&	11.4	&	0.7\\
NGC 660  	&	141.6	&	-47.4	&	15	&	0.9\\
NGC 2146  	&	135.7	&	24.9	&	16.3	&	2.6\\
NGC 3079  	&	157.8	&	48.4	&	17.4	&	2.1\\
NGC 1068 	&	172.1	&	-51.9	&	17.9	&	12.1\\
NGC 1365  	&	238	&	-54.6	&	22.3	&	1.3	\\
Arp 299  	&	141.9	&	55.4	&	46	&	1.6\\
Arp 220  	&	36.6	&	53	&	80	&	0.8	\\
NGC 6240  	&	20.7	&	27.3	&	105	&	1	\\
Mkn 231  	&	121.6	&	60.2	&	183	&	0.8
\enddata
\tablenotetext{1}{Names of SBGs.}
\tablenotetext{2}{Directions of SBGs (galactic coordinates).}
\tablenotetext{3}{Distances from the earth.}
\tablenotetext{4}{Relative flux contributions normalized by a radio flux at 1.4 GHz.}
\end{deluxetable}
An example of the SBG model $F_{\rm org}({\bf n},\theta=10{\,\rm deg})$ is presented in Figure \ref{fig:f_g}. 
As expected from Table \ref{table:aab2018tab1} and Figure \ref{fig:sbg}, a small number  ($<10$) of sources dominates the distribution.
The normalized CR flux model $F_{\rm norm}$ is defined as the weighted sum of the SBG model $F_{\rm org}$ and the isotropic flux model $F_{\rm iso}$: 
\begin{eqnarray}\label{eq:fmod}
\begin{split}
F_{\rm norm}({\bf n},f_{\rm ani},\theta)=f_{\rm ani}F_{\rm org}^{'}({\bf n}, \theta)+(1-f_{\rm ani})F_{\rm iso} \\
F_{\rm org}^{'}({\bf n},\theta)=\frac{F_{\rm org}({\bf n},\theta)}{\int_{4\pi}F_{\rm org}d\Omega}, F_{\rm iso}=1/4\pi.
\end{split}
\end{eqnarray} 

\subsection{Flux mapping}\label{sec:mod}
At this stage, the flux model does not include the GMF deflections. To include them we use a back-propagation technique using the CR propagation code CRPropa3 \citep{Batista2016}.
We adopt the JF12 model \citep{Jansson2012a, Jansson2012b} as the GMF model.
We calculate the trajectories of antiprotons of energy $E$ emitted from the earth to a sphere of $20\,\rm kpc$ radius from the Galactic center (GC) (the galaxy sphere). 
The position of the Earth is defined to be 8.5\, kpc away from the GC, following the JF12 model.
These trajectories represent the trajectories of particles with the same rigidity which can arrive on the Earth through the Galactic sphere. 
The trajectory of a heavier-mass particle with the charge $Ze$ is replaced with the trajectory of a proton with rigidity $R = E/Ze$. 

Based on the CR trajectories obtained through the back-propagation, we convert the CR flux model on the galaxy sphere to that on Earth through the GMF model (flux mapping). 
\begin{figure*}
\centering
\includegraphics[width=1\linewidth]{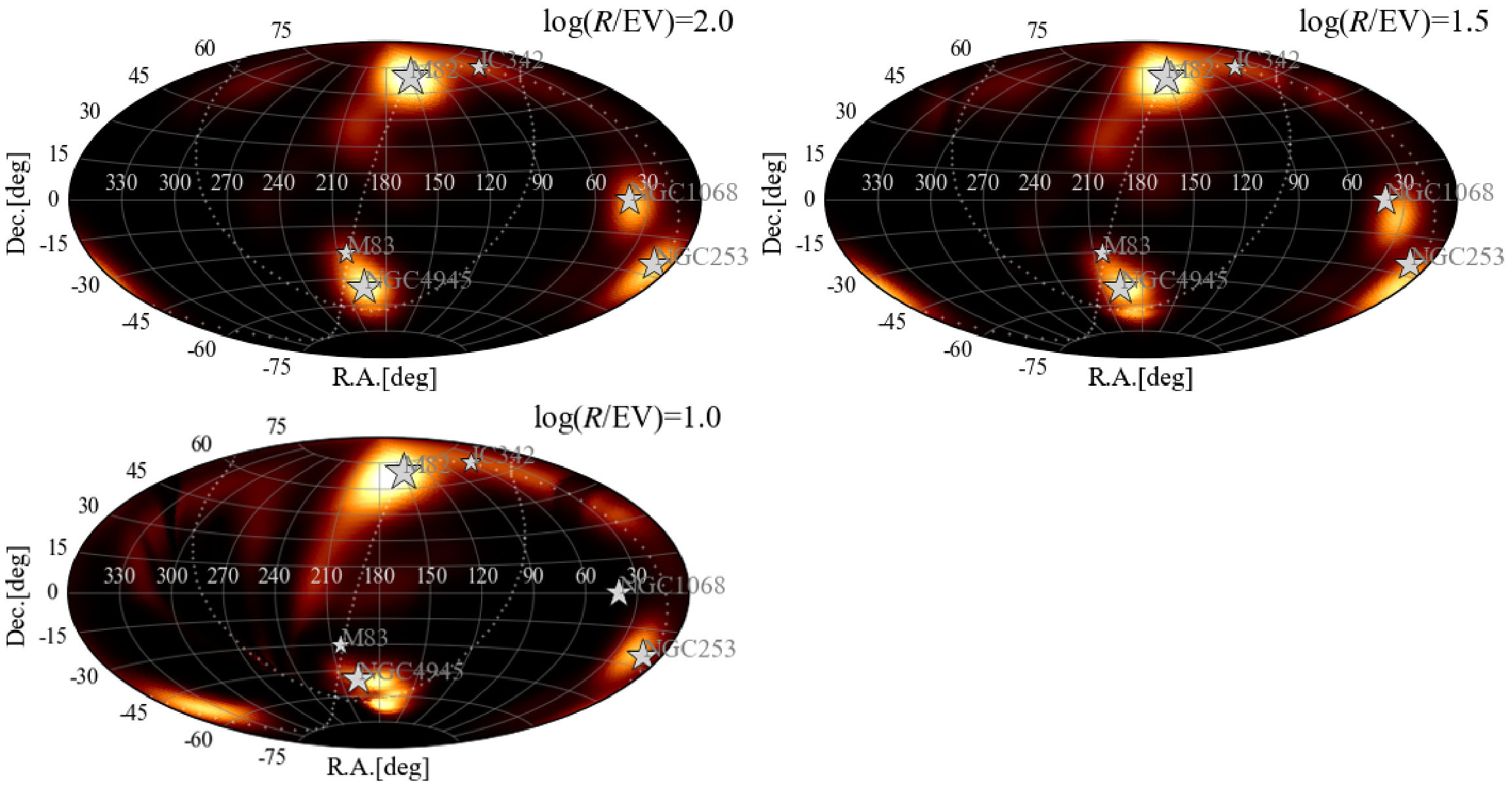}
\caption{
Examples of the SBG model as seen from Earth when $R=10^{2.0},10^{1.5}$ and $10^{1.0}\,\rm EV$ and $\theta =10 \,\deg$ (the JF12 model). 
The color scale is the same as that in Figure \ref{fig:f_g}. 
The rigidity $R$ is shown at the top-right in each panel in the log scale. 
}\label{fig:sub10}
\end{figure*}
We define the original CR flux model as $ F_{\rm org}({\bf n_{\rm org}},\theta)$, where ${\bf n_{\rm org}}$ indicates the direction on the galaxy sphere.
The conversion from the directions on the earth ${\bf n_{\rm earth}}$ to those on the galaxy sphere ${\bf n_{\rm org}}$ of particles with rigidity $R$ is expressed as 
\begin{eqnarray}\label{eq:f_g1}
{\bf n_{\rm org}}=A_{\rm BT}({\bf n_{\rm earth}}, R),
\end{eqnarray}  
where $A_{\rm BT}$ indicates the conversion function.
As Liouville's theorem tells that the flux value along each CR trajectory remains constant \citep{Bradt2008},
we can determine the CR flux on Earth $F_{\rm earth}$ as
\begin{equation}\label{eq:f_e1}
\begin{split}
F_{\rm earth}({\bf n_{\rm earth}},\theta,R)&=F_{\rm org}({\bf n_{\rm org}},\theta)\\
&=F_{\rm org}(A_{\rm BT}({\bf n_{\rm earth}},R),\theta).
\end{split}
\end{equation} 
Because of the nearby source contributions, the photo-nuclear interaction during intergalactic propagation is not taken into account in this study.
Some examples of $F_{\rm earth}$ for different $R$ based on $F_{\rm org}$ in Figure \ref{fig:f_g} are shown in Figure \ref{fig:sub10}.
As shown in Figure \ref{fig:sub10} (top-left), at the highest rigidity ($R=100 \rm \, EV$), the GMF does not affect the CR flux. As the rigidity $R$ becomes lower, the peaks of CR flux around NGC1068 and NGC253 become displaced from the true source directions (Figure \ref{fig:sub10} (top-right) for $\log (R/ \rm EV)=1.5$). At lower rigidity (Figure \ref{fig:sub10} (bottom-left) for $R=10 \rm \, EV$), the displacements become larger as well as the peak around NGC4945 splits along the GP.
Through visual inspections, it is clear that the GMF bias is stronger in the southern hemisphere. 

\subsection{Generation of mock datasets}\label{sec:genmmc}
\begin{figure}
\epsscale{0.7}
\plotone{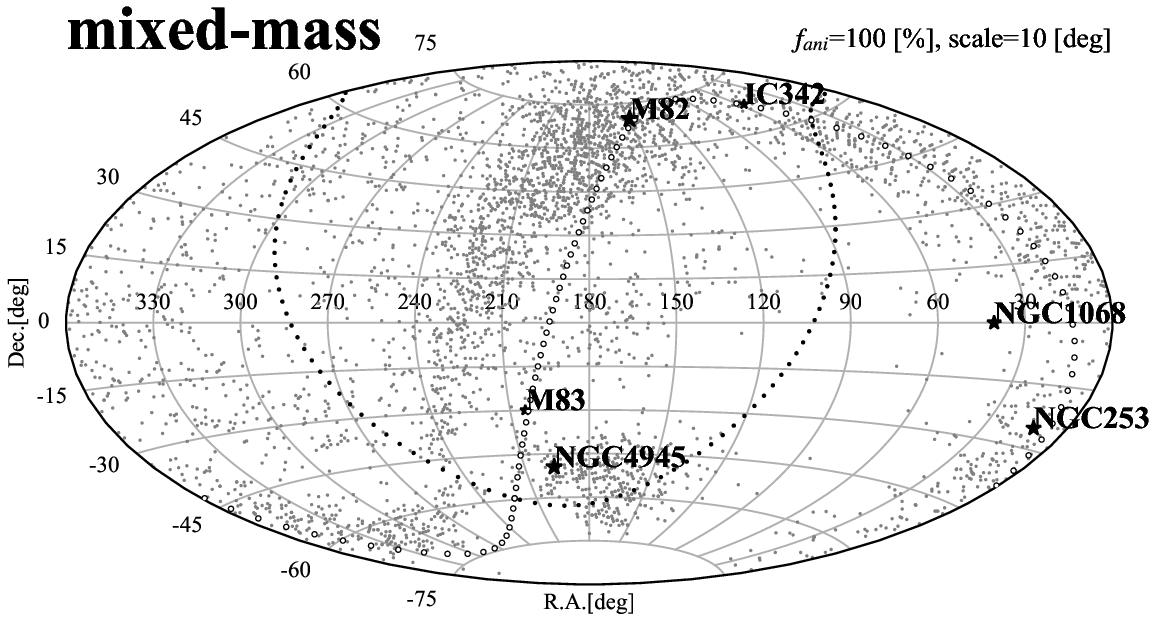}
\caption{
Example of the distribution of the mock events ($(f_{\rm ani}^{\rm true},\theta^{\rm true})=(100\,\%,10\,\deg)$). 
The gray dots show the arrival directions of 4000 mock events.
The directions of the SBGs whose contribution is above 5$\,\%$ are shown by black stars and their name. The area of each star indicates its relative contribution in Tabel \ref{table:aab2018tab1}. Black dots (circles) present GP (SGP).
}
\label{fig:disMMCv3}
\end{figure}
To quantitatively discuss the GMF biases in Section \ref{sec:method}, we generate mock events as follows.
We adopt a best-fit function and parameters given in \cite{Heinze2019} based on the observed UHECRs from the Auger experiment \citep{Aab2017a}. 
In \cite{Heinze2019}, the energy spectrum of each mass ($A$) at the source is assumed by the following function $J_A$, and fitting was performed for the mass composition observed on Earth.
\begin{eqnarray}
\label{eq:CF}
J_{A}(E)={\mathcal J_{A}f_{\rm cut}}(E,Z_{A},R_{\rm max})n_{\rm evol}(z)\left(  \frac{E}{10^9\: \rm GeV} \right)^{-\gamma}
\end{eqnarray}
The cutoff function $f_{\rm cut}$ is given as 
\begin{eqnarray}
\label{eq:heinze5}
f_{\rm cut}=
	\begin{cases}
	1 & (E < Z_A R_{\rm max}) \\
	\exp{\left( 1-\frac{E}{Z_A R_{\rm max}}\right)}  & (E > Z_A R_{\rm max}).
	\end{cases}
\end{eqnarray}
Because we only focus on the nearby sources, the redshift evolution term $n_{\rm evol}(z)$ is approximated to be $1$.
We also assume that the mass composition observed on Earth and that at the source are the same.
The fractions of elements are defined as $f_A$=$\mathcal J_A/\Sigma_A \mathcal J_A$ at 10 EeV in \cite{Heinze2019}. 
We adapt the best-fit parameters from \cite{Heinze2019} as $\gamma=-0.80$ and $R_{\rm max}=1.6\,\rm EV$. 
We also adapt the values of $f_A$ as $(\rm ^{1} H,\:^{4}He,\:^{14}N,\;^{28}Si,\:^{56}Fe)=(0.0,82.0,17.3,0.6,2.0\cdot 10^{-2}) [\%]$. 
According to this mass fraction and spectra, we determine the mass and energy, {\it i.e.} $R$, of each mock event.
The energy $E$ of a mock event is randomly sampled from the spectrum and the rigidity $R$ of the event is calculated through the formula $R=E/Ze$. 
We adapt the minimum energy $E_{\rm min}= 40\rm \, EeV$ according to the previous studies \citep[$E_{\rm min}=39 \,\rm EeV$]{Aab2018}. 
Using the selected rigidity $R$, we determine the arrival direction of the event based on the CR flux model defined in Section \ref{sec:mod}.

An example of the distribution of the mock event arrival directions is provided in Figure \ref{fig:disMMCv3}.
The distribution in Figure \ref{fig:disMMCv3} is similar to the distribution of pure-carbon case (Figure \ref{fig:dispure} in Appendix \ref{sec:mockpure}). Although we can see clusterings around M82 and NGC4945, the centers of the distributions are displaced from the source directions. The events that originated from NGC1068 and NGC253 are mostly deflected. 
This suggests that the real UHECR distribution cannot be reproduced with a single isotropic smearing, and the deflections by the GMF depend on the arrival directions.

\subsection{The sky coverage of experiments}\label{sec:exp}
To make the comparison with the analysis of the observed UHECRs \citep{Aab2018, Abbasi2018}, the sky coverage of the TA and Auger experiments is considered based on equations given by \cite{Sommers2001}. 
In \cite{Sommers2001}, the sky coverage $\omega({\bf n}_{\rm CR})$ depends on the declination $\delta$: 
\begin{eqnarray}
\label{eq:sommer}
\omega(\delta) \propto {\rm cos}(a_0){\rm cos}(\delta){\rm sin}(\alpha_{\rm m})+\alpha_{\rm m} {\rm sin}(a_0){\rm sin}(\delta). 
\end{eqnarray}
\begin{align}
\alpha_{\rm m}=
\begin{cases}
    0\: & (\xi>1) \\
    \pi\: & (\xi<-1) \\
    {\rm cos}^{-1}(\xi)\: & (-1<\xi<1)
\end{cases}
\end{align}
\begin{eqnarray}
\xi=\frac{ {\rm cos}(\theta_{\rm m})-{\rm sin}(a_0){\rm sin}(\delta)}{{\rm cos}(a_0){\rm cos(\delta)}}, 
\end{eqnarray}
where $\theta_{\rm m}$ is the maximum zenith angle and $a_0$ is the latitude of the experimental site. 
We adopt the latitude $a_{0}=39.3\,\deg$ ($-35.2\,\deg$) and maximum zenith angle $\theta _{\rm m}=55\,\deg$ ($60\,\deg$) for TA (Auger) experiment. 
From the all-sky datasets, we randomly select mock events with the probability of each sky coverage. 
Out of the 4000 mock events in each dataset, approximately $1000$ mock events are selected by each of the TA and Auger coverage. 
We define the dataset selected by the sky coverage of TA (Auger) as the north-sky (south-sky) dataset. 

\section{Analysis} \label{sec:method}
In order to investigate how much GMF deflections affects the estimated parameters, we conduct the same maximum-likelihood analysis in \cite{Aab2018} to the mock datasets.
We test two hypotheses for the CR flux models.
One is a flux with non-zero $f_{\rm ani}$ with the SBG model ($F_{\rm norm}$) and the other is the isotropic flux ($F_{\rm iso}$), {\it i.e.}, $f_{\rm ani}=0$.
The test statistics $TS$ are calculated as a log-likelihood ratio: 
\begin{eqnarray}
\label{eq:ts}
TS=2\ln (L(F_{\rm norm})/L(F_{\rm iso})). 
\end{eqnarray} 
A likelihood of each model $L(F)$ is given as
\begin{eqnarray}
\label{eq:L}
L(F)=\prod_{\rm CR} \frac{F({\bf n}_{\rm CR})\omega ({\bf n}_{\rm CR})}{\int_{4\pi}F({\bf n})\omega ({\bf n})d\Omega},
\end{eqnarray} 
where $F$, $\omega({\bf n}_{\rm CR})$, and ${\bf n}_{\rm CR}$ are the normalized CR flux model, the sky coverage of each experiment (Equations \ref{eq:sommer}), and the arrival directions of observed UHECRs, respectively.

By scanning the set of parameters $(f_{\rm ani},\theta)$, the best-fit parameters that maximize the $TS$ in Equation \ref{eq:ts} are determined. 

\section{Results}\label{sec:difMMCv3}
\begin{figure*}
	\centering
	\includegraphics[width=1\linewidth]{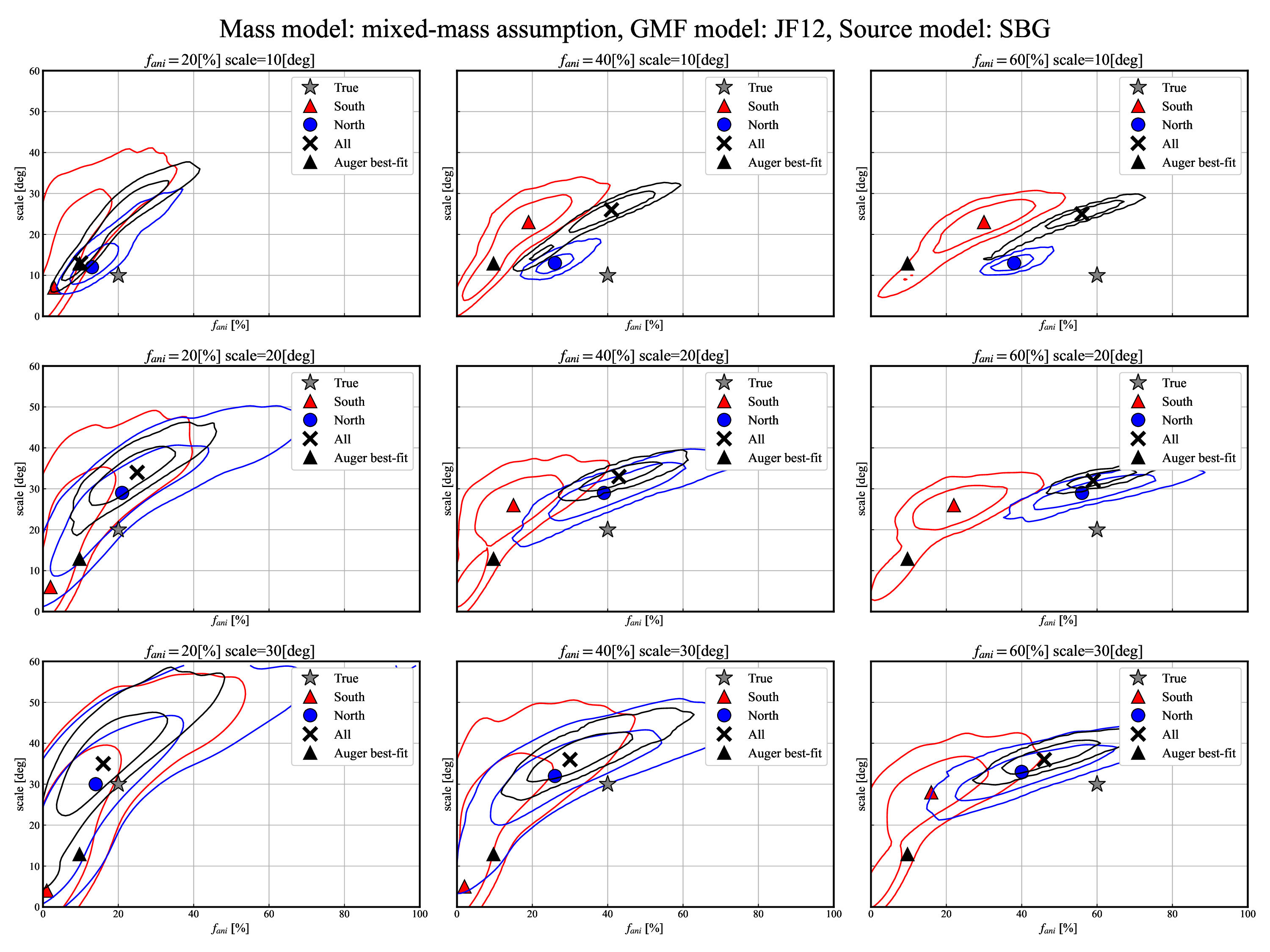}
\caption{
Distributions of the best-fit parameters for the 1000 mock event datasets. 
From the left to right column, the true parameter $f_{\rm ani}^{\rm true}$ is given as $20\,\%$, $40\,\%$ and $60\,\%$. 
From the top to the bottom row, the true parameter $\theta$ is 10, 20, and 30$\,\deg$.
The true parameters are marked by the grey stars.
The black, blue, and red contours indicate the 68$\,\%$ and 95$\,\%$ tile containment for all, north and south-sky datasets, respectively. 
The black cross, blue circle, and red triangle show the most frequent values $(\tilde{f}_{\rm ani},\tilde{\theta})$ for all, north and south-sky datasets, respectively. 
The distributions of best-fit parameters are smoothed with a kernel-Gaussian distribution. }
The best-fit parameter $(f_{\rm ani}^{\rm Auger},\theta^{\rm Auger})=(9.7 \,\%,12.9 \,\rm deg)$ in \cite{Aab2018} is shown as a black triangle.
\label{fig:difMMCv3}
\end{figure*}
We show the distribution of the best-fit parameters for 1000 mock event datasets in Figures \ref{fig:difMMCv3}.
The different panels show the results for the different true parameters $(f_{\rm ani}^{\rm true},\theta^{\rm true})$ indicated at the top of each panel and by the gray star at each panel. 
From the left to right column, the true parameter $f_{\rm ani}^{\rm true}$ is given as $20\,\%$, $40\,\%$ and $60\,\%$.
From the top to the bottom row, the true parameter $\theta$ is given as 10, 20, and 30$\,\deg$.
The black, blue, and red contours indicate the 68$\,\%$ and 95$\,\%$ tile containment for the distributions for all-sky, north-sky, and south-sky datasets, respectively. 
The black cross, blue circle, and red triangle show the most frequent values $(\tilde{f}_{\rm ani},\tilde{\theta})$ of the best-fit parameters of all-sky, north-sky, and south-sky datasets, respectively.

The distributions of all-sky, north-sky, and south-sky datasets do not agree with each other, especially for a higher anisotropic fraction $f_{\rm ani}^{\rm true}$ (the right panels).
When the anisotropic fraction $f_{\rm ani}^{\rm true}$ is larger and separation angular scale $\theta^{\rm true}$ is smaller (4 upper-right panels), the estimated separation angular scale $\theta$ becomes larger due to the deflection of the GMF. When the anisotropic fraction $f_{\rm ani}^{\rm true}$ is smaller (left panels), the distribution of all, north, and south-sky datasets are similar due to the low contrast between the SBG model and the isotropic backgrounds.

Focusing on the north-south difference, the GMF affects the results of south-sky datasets more than north-sky datasets. 
This can be explained by two reasons: first, the GMF deflections are larger around GC. In the rest of the sky, the GMF deflection is larger in the Galactic-south (see also Figure 11 in \cite{Jansson2012a}).
NGC253 and NGC1068, which contribute to the south-sky datasets, are located in the region of the sky where the GMF deflections are large.
In any parameter sets $(f_{\rm ani}^{\rm true},\theta^{\rm true})$, it is found that the most frequent value of anisotropic fraction $\tilde{f}_{\rm ani}$ for the south sky datasets (red triangles) are largely underestimated, namely below 50$\%$ of the true ones $f_{\rm ani}^{\rm true}$ (grey stars). 
Regardless of the true parameters $(f_{\rm ani}^{\rm true},\theta^{\rm true})$, the distributions for the south-sky datasets include the best-fit parameters $(f_{\rm ani}^{\rm Auger},\theta^{\rm Auger})=(9.7 \,\%,12.9 \,\rm deg)$ indicated by the black triangles in 2$\sigma$ contour. We discuss this tendency in Section \ref{sec:bestfit}. 

\section{Discussions}
\subsection{The uncertainty of the GMF models}\label{sec:unGMF}
In this section, we discuss the effect of uncertainty in the GMF models. 
To test the effects caused by the uncertainty of the halo components of the JF12 model, we conduct the same analysis but change the halo components in the model within $1\sigma$ uncertainty, generate the mock event datasets, and repeat the analysis.
It is found that the uncertainty of the halo components does not have a large effect on the most frequent parameters $(\tilde{f}_{\rm ani},\tilde{\theta})$. 

For an independent comparison with the JF12 model, we also refer to the  Pshirkov $\&$ Tinyakov 2011 model (PT11) \citep{Pshirkov2011}. 
We generate the mock event datasets based on the PT11 model.
Except for the GMF model, the other assumptions (the SBG model and mixed-mass composition) are the same. 
Although the separation angular scale $\theta$ in the south-sky datasets becomes smaller than with the JF12 model, the most frequent values of anisotropic fractions $\tilde{f}_{\rm ani}$ are also reduced by more than $50\,\%$ compared to the true value $f_{\rm ani}^{\rm true}$ (see also Figure \ref{fig:difPT11MMCv3} in Appendix \ref{app:A1}). 

\subsection{Comparison with the best-fit parameters in \cite{Aab2018}}\label{sec:bestfit}
\begin{figure}
\centering
\includegraphics[width=1\linewidth]{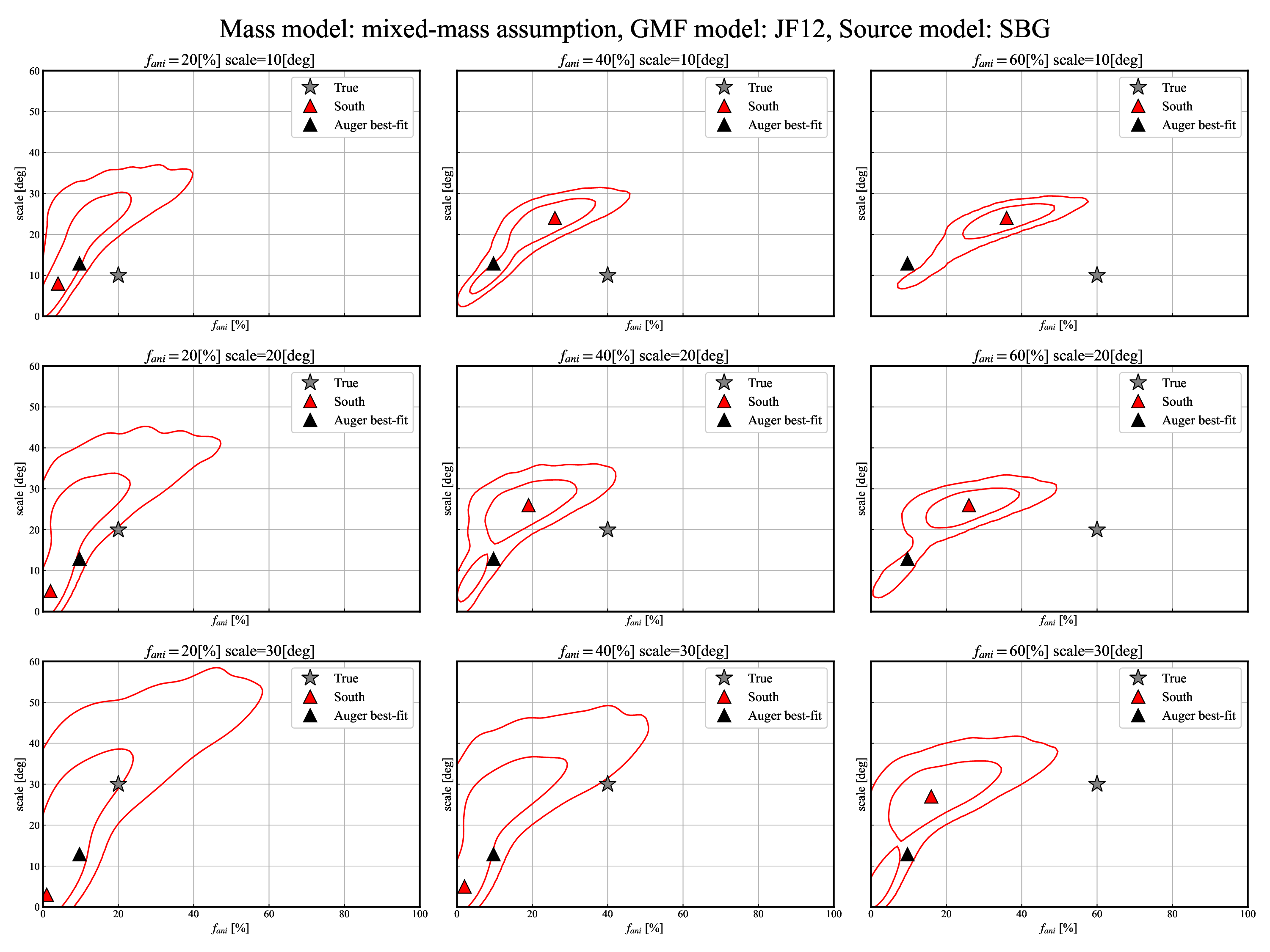}
\caption{
Examples of the distributions of best-fit parameters estimated from the mock event datasets. 
Red contours show 68 and 95$\%$ tile containments of the best-fit parameters. 
The red triangle shows the most frequent value of the best-fit parameters $(\tilde{f}_{\rm ani},\tilde{\theta})$ of the mock event datasets. 
The best-fit parameter $(f_{\rm ani}^{\rm Auger},\theta^{\rm Auger})=(9.7 \,\%,12.9 \,\rm deg)$ in \cite{Aab2018} is shown as a black triangle.
Note that the mock event datasets are generated with a set of assumptions for the source model (SBG), the GMF model (the JF12), and the mass composition model \citep{Heinze2019}.
	}
	\label{fig:contour}
\end{figure}
\begin{figure}
	\centering
	\includegraphics[width=0.7\linewidth]{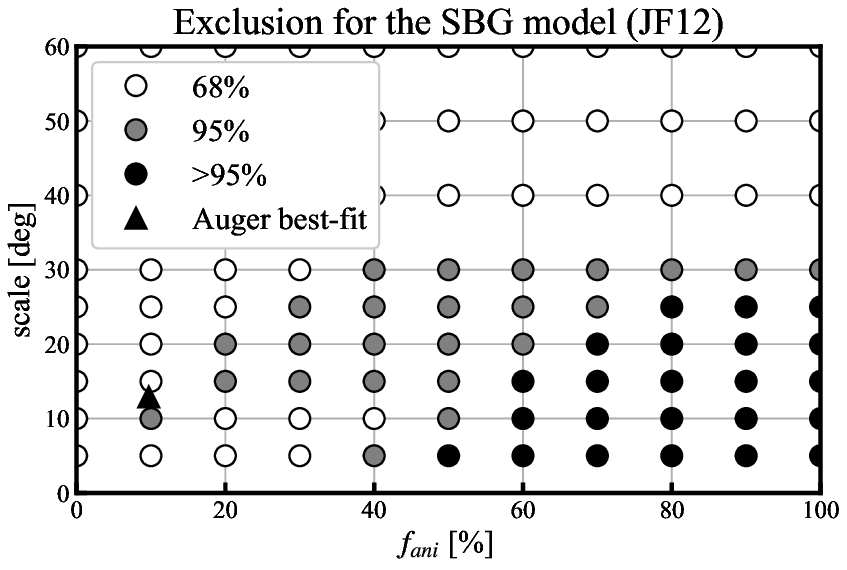}
	\caption{
Excluded region for the best-fit parameter in \cite{Aab2018}.
Circles show the searched true parameters $(f_{\rm ani}^{\rm true},\theta^{\rm true})$. 
White (gray) color indicates} the true parameters $(f_{\rm ani}^{\rm true},\theta^{\rm true})$ which reproduce the best-fit parameter $(f_{\rm ani}^{\rm Auger},\theta^{\rm Auger})=(9.7 \,\%,12.9 \,\rm deg)$ in \cite{Aab2018} within 68 (95) percentile. 
The best-fit parameter $(f_{\rm ani}^{\rm Auger},\theta^{\rm Auger})=(9.7 \,\%,12.9 \,\rm deg)$ in \cite{Aab2018} is shown as a black triangle. 
The parameters are scanned with resolutions of $\Delta f_{\rm ani}^{\rm true}=10\,\%$ and $\Delta \theta^{\rm true}=5\,\deg$.
	\label{fig:tile}
\end{figure}
In this section, we search a set of the true parameters $(f_{\rm ani}^{\rm true},\theta^{\rm true})$ that is compatible with the best-fit parameters $(f_{\rm ani}^{\rm Auger},\theta^{\rm Auger})$. 
We generate 4000 south-sky datasets in the same manner as in Section \ref{sec:genmmc}, but each has the same number of events (894 events) used in the analysis of \cite{Aab2018}.
These mock event datasets are analyzed in the same manner and the best-fit parameters are obtained.
Examples of the distributions of 4000 best-fit parameters are shown in Figure \ref{fig:contour}.
Because of the GMF bias and the statistical fluctuation regardless of the true parameters $(f_{\rm ani}^{\rm true},\theta^{\rm true})$ marked by the grey star, the estimated parameters tend to distribute around the Auger best-fit parameters $(f_{\rm ani}^{\rm Auger},\theta^{\rm Auger})$ marked by the black triangle within 68\% or 95\% containment levels.
We classify the true parameters $(f_{\rm ani}^{\rm true},\theta^{\rm true})$ according to whether $(f_{\rm ani}^{\rm Auger},\theta^{\rm Auger})$ is contained in the $68\,\%$ or $95\,\%$ contours. 
Figure \ref{fig:tile} shows the result of this classification. 
From Figure \ref{fig:tile}, except in the right-bottom corner, a wide range of parameters is still compatible with $(f_{\rm ani}^{\rm Auger},\theta^{\rm Auger})$.
Considering the GMF effect and the mass-dependent energy spectrum, a large contribution of SBGs to the UHECR flux is still possible.

\subsection{Maximum likelihood analysis method with the CR flux models on the earth}\label{sec:newLA}
For the calculations of likelihood (Equation \ref{eq:L}) and $TS$ (Equation \ref{eq:ts}), we use the original CR flux model $F_{\rm org}$ instead of the CR flux model on the earth $F_{\rm earth}$ (Equation \ref{eq:fmod}).
This is what causes the GMF bias in the parameter estimations. 
To reduce the GMF bias in the previous parameter estimation, it is necessary to replace $F_{\rm org}$ with $F_{\rm earth}$ in Equation \ref{eq:fmod}.
Note that this analysis is valid only when the GMF and mass-dependent spectrum models are correct.
In other words, we need to test a set of assumptions together. 
We rewrite Equation \ref{eq:fmod} as follows: 
\begin{equation}\label{eq:fearth}
\begin{split}
F_{\rm norm}({\bf n},f_{\rm ani},\theta,R)=f_{\rm ani}F_{\rm earth}^{'}({\bf n},\theta,R)+(1-f_{\rm ani})F_{\rm iso} \\
F_{\rm earth}^{'}=\frac{F_{\rm earth}({\bf n},\theta,R)}{\int_{4\pi}F_{\rm earth}d\Omega}, \: F_{\rm iso}=1/4\pi
\end{split}
\end{equation} 
Here, $F_{\rm earth}({\bf n_{\rm earth}},\theta,R)$ is obtained using Equation \ref{eq:f_e1}. 
Thus, we can rewrite the CR flux models from the sources $F_{\rm earth}^{'}({\bf n},f_{\rm ani},\theta, R)$ as
\begin{equation}
\label{eq:fnorm2}
F_{\rm earth}^{'}({\bf n}_{\rm CR},f_{\rm ani},\theta,R_{\rm CR})=\frac{F_{\rm org}(A_{\rm BT}({\bf n}_{\rm CR}, R_{\rm CR}),\theta)}{\int_{4\pi}F_{\rm org}(A_{\rm BT}({\bf n}, R),\theta)d\Omega}.
\end{equation} 
The denominator $\int_{4\pi}F_{\rm org}(A_{\rm BT}({\bf n}, R),\theta)d\Omega$ in Equation \ref{eq:fnorm2} is derived by integrating $F_{\rm earth}$. 

\begin{figure}[h]
	\centering
	\includegraphics[width=1\linewidth]{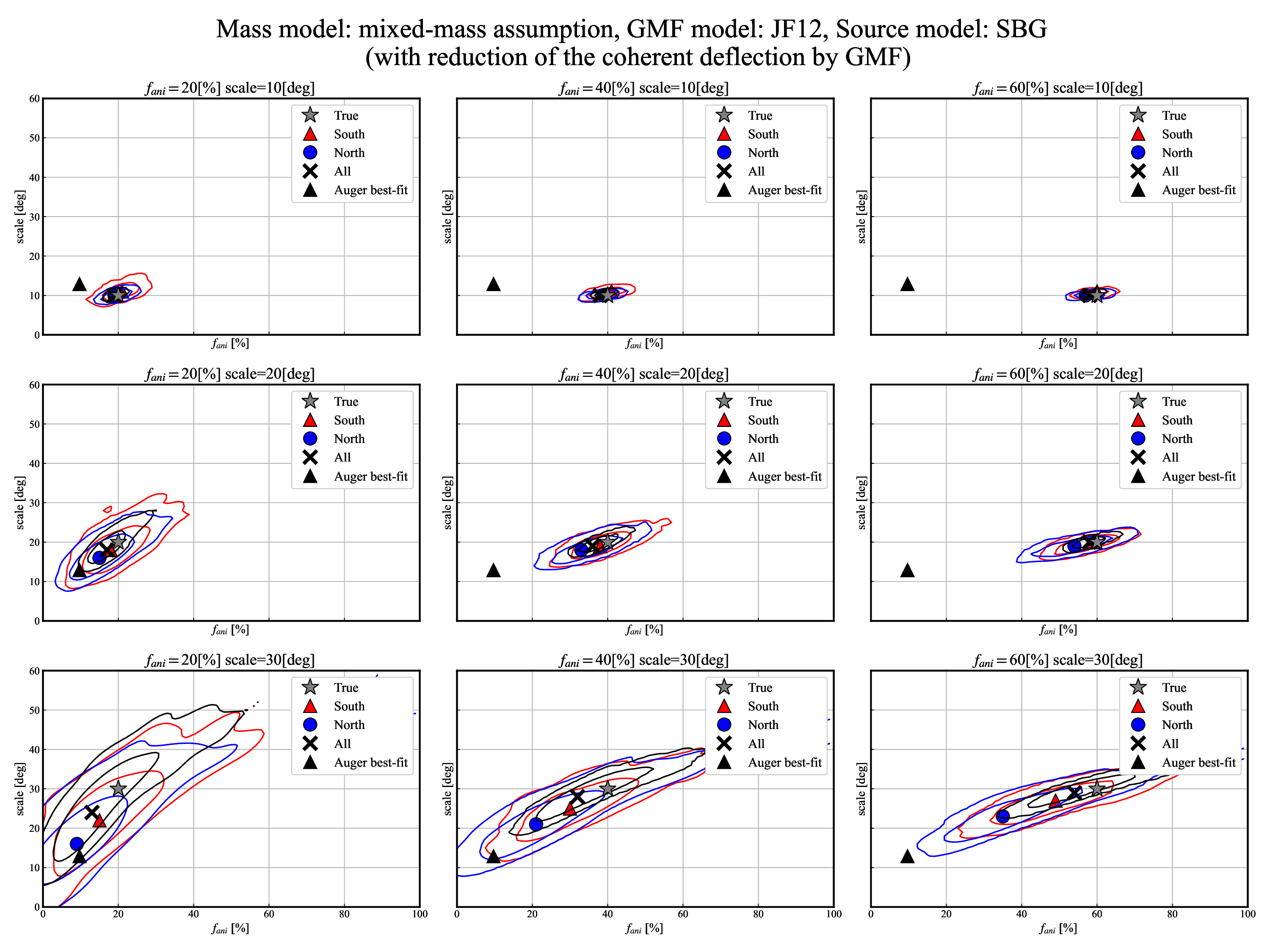}
	\caption{
Same as Figure \ref{fig:difMMCv3} but with the improved analysis method in Section \ref{sec:newLA}.
The analysis is applied for the same mock event datasets with mixed-mass assumption and JF12 model in Figure \ref{fig:difMMCv3}. 
	}
	\label{fig:fix_MMCv3}
\end{figure}
We conduct the improved maximum-likelihood analysis following Equation \ref{eq:fearth} to the same datasets as in Sections \ref{sec:difMMCv3}.
Note that the new analysis is carried out assuming that we know the event-by-event mass of each mock event. 
Figure \ref{fig:fix_MMCv3} illustrates the results in the same manner as Figure \ref{fig:difMMCv3}, but for the estimation following Equation \ref{eq:fearth}. 
In all cases, the analysis improves the estimates of the parameters $(f_{\rm ani},\theta)$. 
Specifically, the GMF bias is reduced when the separation angular scales $\theta$ are small. 
For a larger separation angular scale, there is still a significant difference between the estimated parameters and the true parameters $(f_{\rm ani}^{\rm true},\theta^{\rm true})$.
Because the GMF bias caused by the regular component of the GMF is effectively reduced within $1\sigma$ contour, the origins of the dispersion should be the statistical fluctuation.
Future observations with large statistics are expected to reduce the dispersion.
In this analysis, we assumed a perfect event-by-event energy and (rigidity) resolution, but the effect of realistic resolutions will be discussed in a future publication. 

\subsection{Discussions on the random GMF and EGMF}\label{sec:random}
In this section, we refer to components of magnetic fields which are not fully considered in this study.
To reveal the effects of coherent deflection by the GMF, we only focus on the regular component of the GMF.
We assume that the separation angular scale $\theta$ includes both the EGMF and a random component of the GMF (random GMF). 
\cite{Aab2018,Abbasi2018} also includes regular components of the GMF. 
In general, the deflections by the EGMF and the random GMF also should have a rigidity dependence.
\cite{Bray2018} suggests the upper limit to the EGMF at the $\sim\, \rm nG$ scale. 
Although the upper limit to the EGMF is smaller than that of the GMF, the distance between each source and the Earth is much larger than the radius of our Galaxy.
We need to consider this distance dependence (see also \citealt{Anchordoqui2019}).
A random component of the GMF has an arrival direction dependency which is the same as for the regular component of the GMF. 
\cite{Pshirkov2013} investigated the random deflections in the GMF.
They suggested that the random component of the GMF deflects $40 \, \rm EeV$ protons by less than $1$--$2 \, \rm deg$ in most of the sky and $\sim 5\,\deg$ along the GP.

Although the physically correct description of random components of the GMF and EGMF is important, it is out of the scope of this study.
The effect of random components in the GMF and EGMF also should have an arrival-direction dependency and rigidity-dependency.
We will take them into account in a future realistic model.
\section{Summary}
We estimate the biases caused by the coherent deflections due to the Galactic magnetic field in searches for UHECR sources in the literature.
We generated mock event datasets with a set of assumptions for a source model \citep{Aab2018}, coherent deflection by a GMF model \citep{Jansson2012a, Jansson2012b}, and a mass-composition model \citep{Heinze2019}, and conduct maximum-likelihood analysis on the datasets neglecting the GMF in the same manner as in previous studies. 
Our major results are listed below:
\begin{enumerate}
    \item The distributions of the estimated parameters $(f_{\rm ani},\theta)$ are displaced from the true parameters $(f_{\rm ani}^{\rm true},\theta^{\rm true})$.  
    This confirms the existence of the GMF bias.
    \item The distributions of the estimated parameters $(f_{\rm ani},\theta)$ in all-sky, north-sky, and south-sky datasets do not agree with each other.
    The directional or sky dependence of the GMF bias is also confirmed.
    \item We find that the estimated $f_{\rm ani}$ is systematically reduced by more than $50\,\%$ in the south-sky datasets. 
    \item We search for the true parameters $(f_{\rm ani}^{\rm true},\theta^{\rm true})$ that are compatible with the best-fit parameters reported in \cite{Aab2018} taking into account the number of events used in their study. 
    Except for the narrow region with large anisotropic fraction $f_{\rm ani}^{\rm true}$ and small separation angular scale $\theta^{\rm true}$, a wide parameter space is still compatible with the experimental result within $95\,\%$ C. L. 
    \item We develop a maximum-likelihood analysis taking into account the GMF deflections and confirm that the parameters would be correctly estimated within $1\sigma$ contour under the ideal condition that we know the event-by-event energy and mass of each UHECR event and the GMF structure.
\end{enumerate}
Note that again this study is conducted under a specific set of assumptions: the source model \citep{Aab2018}, magnetic field model \citep{Jansson2012a, Jansson2012b}, energy spectrum and mass composition \citep{Tsunesada2017, Heinze2019}. 

Although these models and the assumptions are to be tested and updated regularly, the technique in Section \ref{sec:newLA} can be applied to future models and updated observational datasets. 
The extension of the TA and Auger experiments \citep[TA$\times$4 and AugerPrime]{Abbasi2021, Castellina2019} and next-generation UHECR observation \citep{GCOS2021} will play an important role. 
The improvement of the GMF and CR propagation models also leads us to more realistic source searches \citep{Boulanger2018}.

\begin{acknowledgments}
We thank the members of the Telescope Array collaboration for fruitful discussions.
We are grateful to Peter Tinyakov, Anatoli Fedynitch, and Federico Urban for fruitful discussions and revisions.
This work was supported by JSPS KAKENHI Grant Numbers JP19J11429, JP19KK0074 and the joint research program of the Institute for Cosmic Ray Research (ICRR), the University of Tokyo. 
E.K. is thankful to supports from ``Pioneering Program of RIKEN for Evolution of Matter in the Universe (r-EMU)".
\end{acknowledgments}

\appendix
\section{Analysis with single-mass assumption}\label{app:A1}
\subsection{Mock datasets with a single-mass assumption}\label{sec:mockpure}
\begin{figure}
	\centering
	\includegraphics[width=1\linewidth]{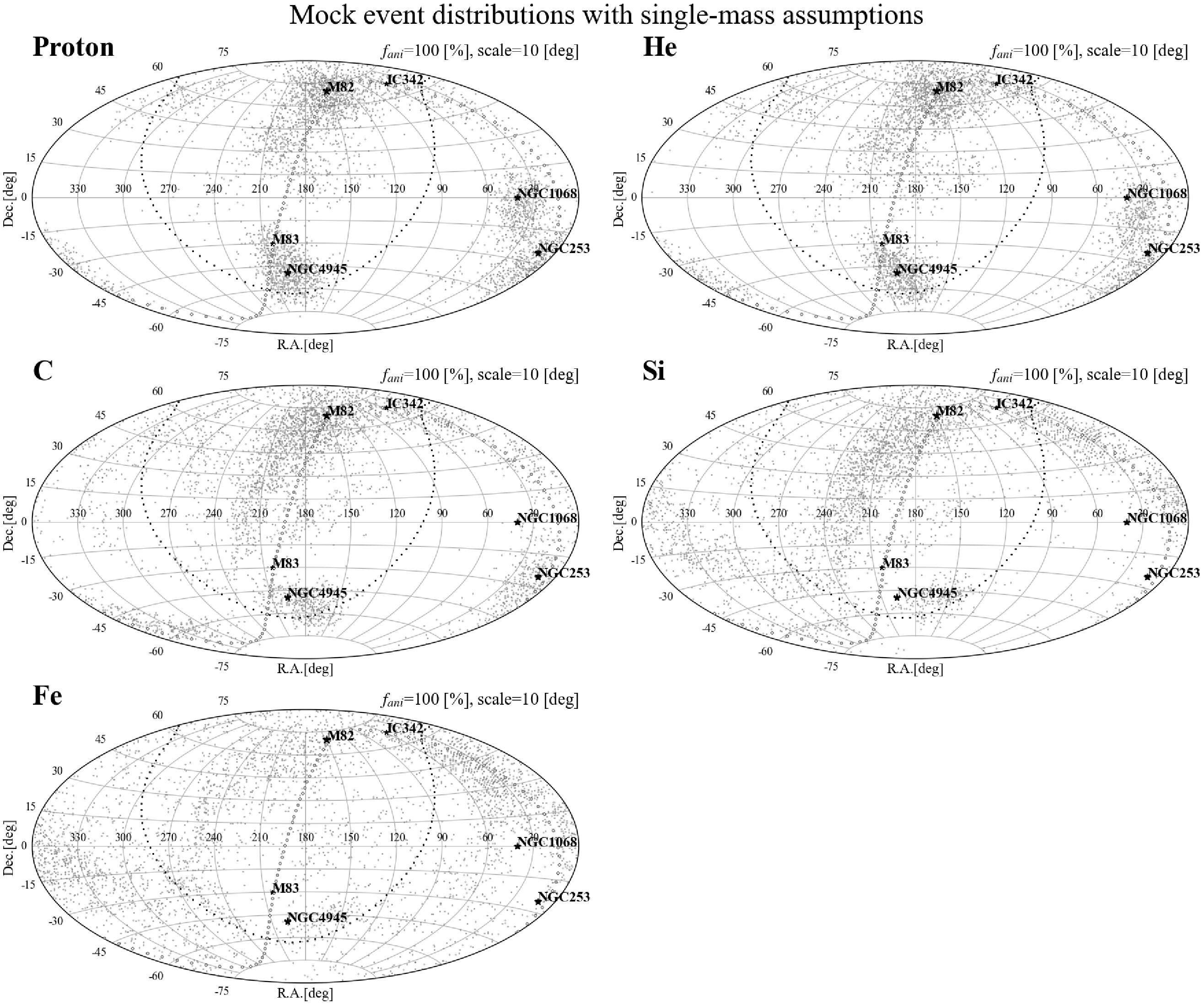}
\caption{
Examples of the distribution of the mock event arrival directions with the single-mass assumptions ($(f_{\rm ani}^{\rm true},\theta^{\rm true}) = (100\,\%, 10 \rm \, deg)$ and 4000 events over all-sky). 
}
\label{fig:dispure}
\end{figure}
As single-mass assumptions, we test the pure-proton, He, C, Si, and Fe cases. 
We fix the energy spectrum as a broken-power law with spectral indexes $\gamma=-2.69\:(E<10^{1.81}\,\rm EeV)$ and $\gamma=-4.63 \:(E>10^{1.81}\,\rm EeV)$, which is as reported by the TA experiment \citep{Tsunesada2017}.
In the same manner as Section \ref{sec:genmmc}, we choose the arrival direction of the anisotropic event based on the generated CR flux models $F_{\rm earth}$.
Examples of the distribution of mock event arrival directions with the single-mass assumption and with $(f_{\rm ani}^{\rm true},\theta^{\rm true}) = (100\,\%, 10 \rm \, deg)$ are shown in Figure \ref{fig:dispure}.
For light masses like proton and Helium, the distributions are similar to that of Figure \ref{fig:f_g}, which means the GMF bias is small.
On the other hand, for the heavier masses, the distortion due to the GMF is significant.
In the pure-Fe assumption, the clusterings of events around the top-4 contributing SBGs (M82, NGC4945, NGC1068, and NGC253) are not seen.

\subsection{Estimated parameters in previous studies with single-mass assumption}\label{sec:difpure}
\begin{figure}
	\centering
	\includegraphics[width=1\linewidth]{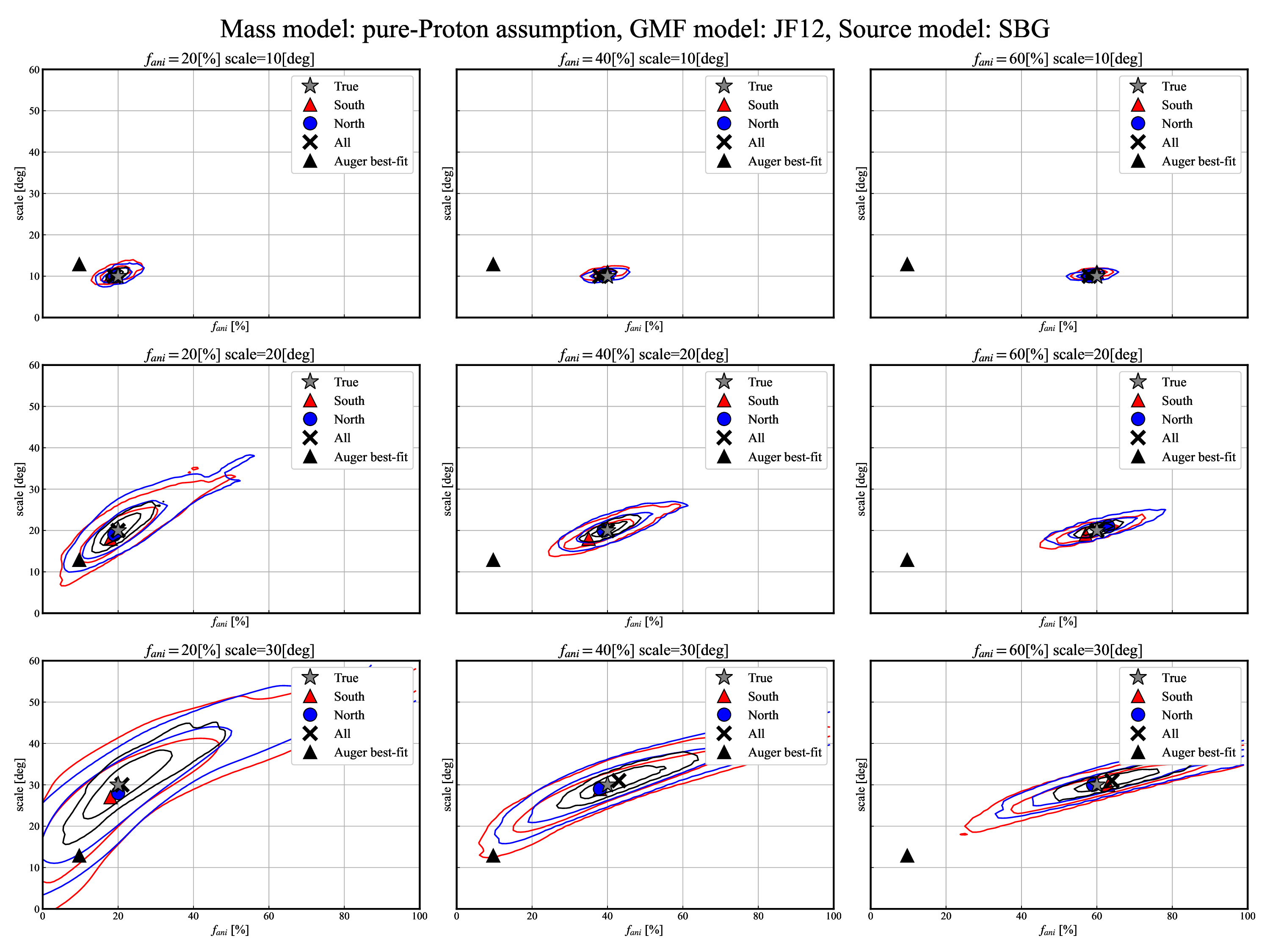}
\caption{
Same as Figure \ref{fig:difMMCv3} but for the pure-proton case.
}
\label{fig:difpureP}
\end{figure}
\begin{figure}
\centering
\includegraphics[width=1\linewidth]{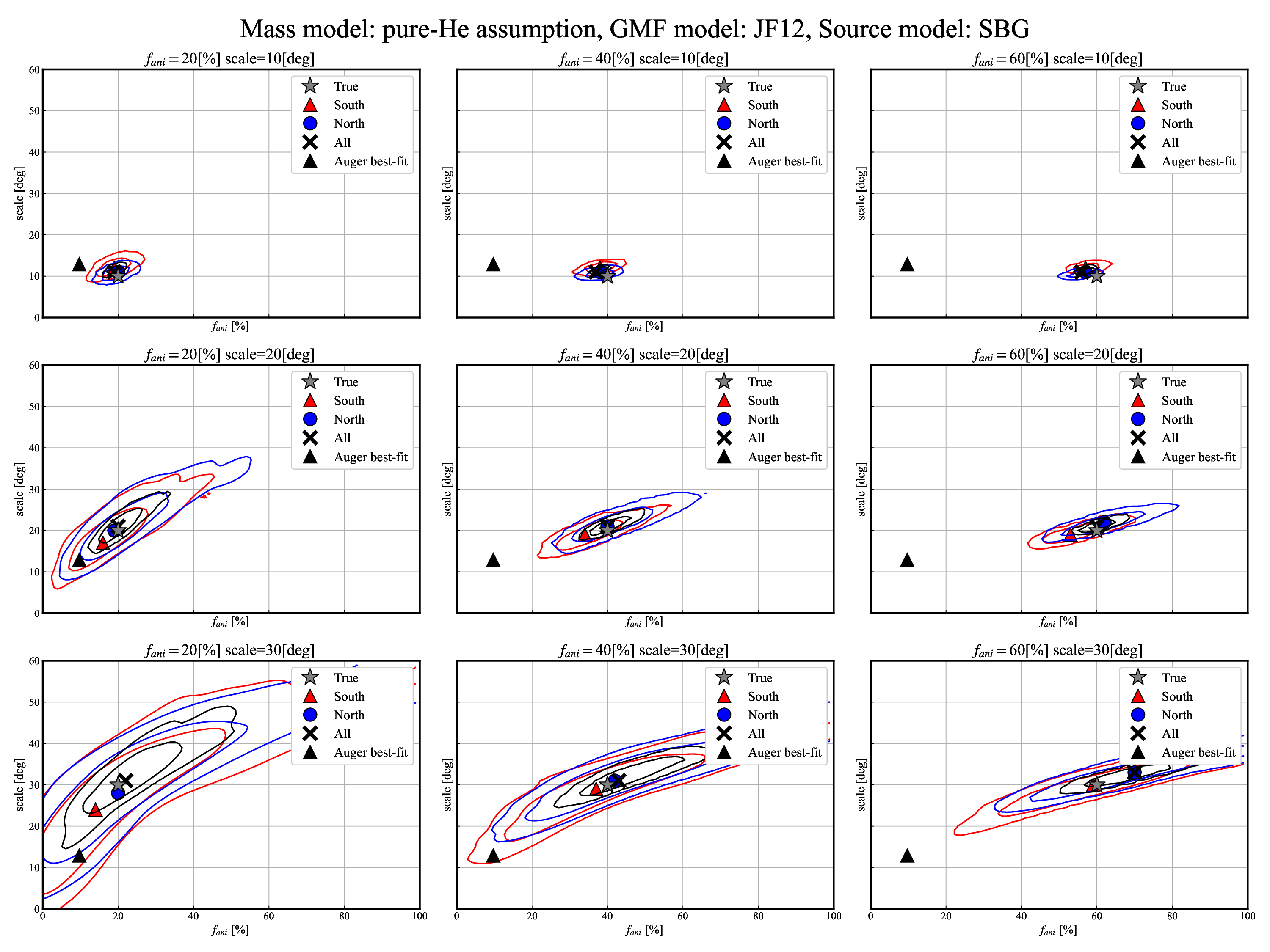}
\caption{
Same as Figure \ref{fig:difpureP} but for the pure-He case. 
}
\label{fig:difpureHe}
\end{figure}
\begin{figure}
\centering
\includegraphics[width=1\linewidth]{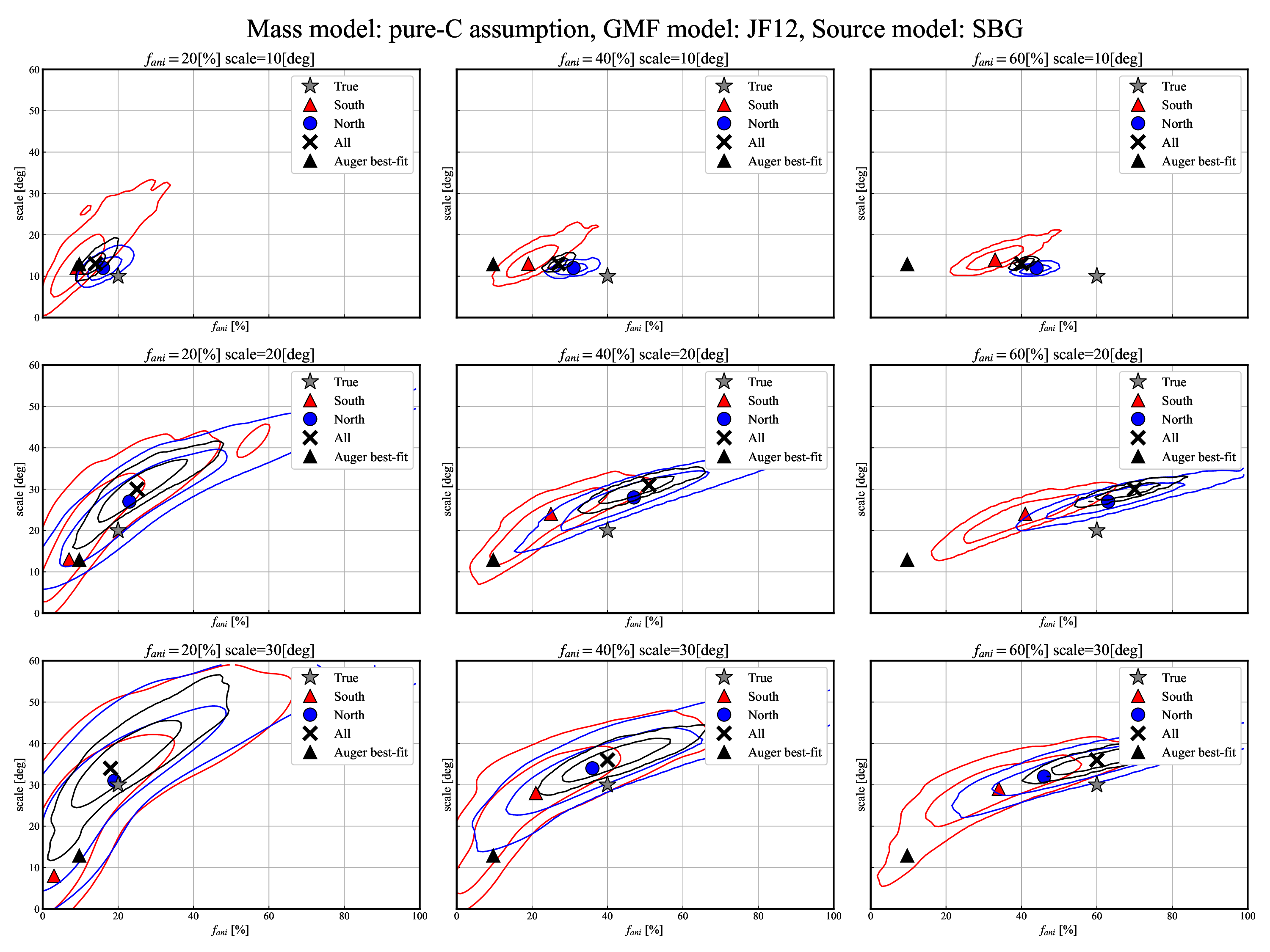}
\caption{
Same as Figure \ref{fig:difpureP} but for the pure-C case. 
}
\label{fig:difpureC}
\end{figure}
\begin{figure}
\centering
\includegraphics[width=1\linewidth]{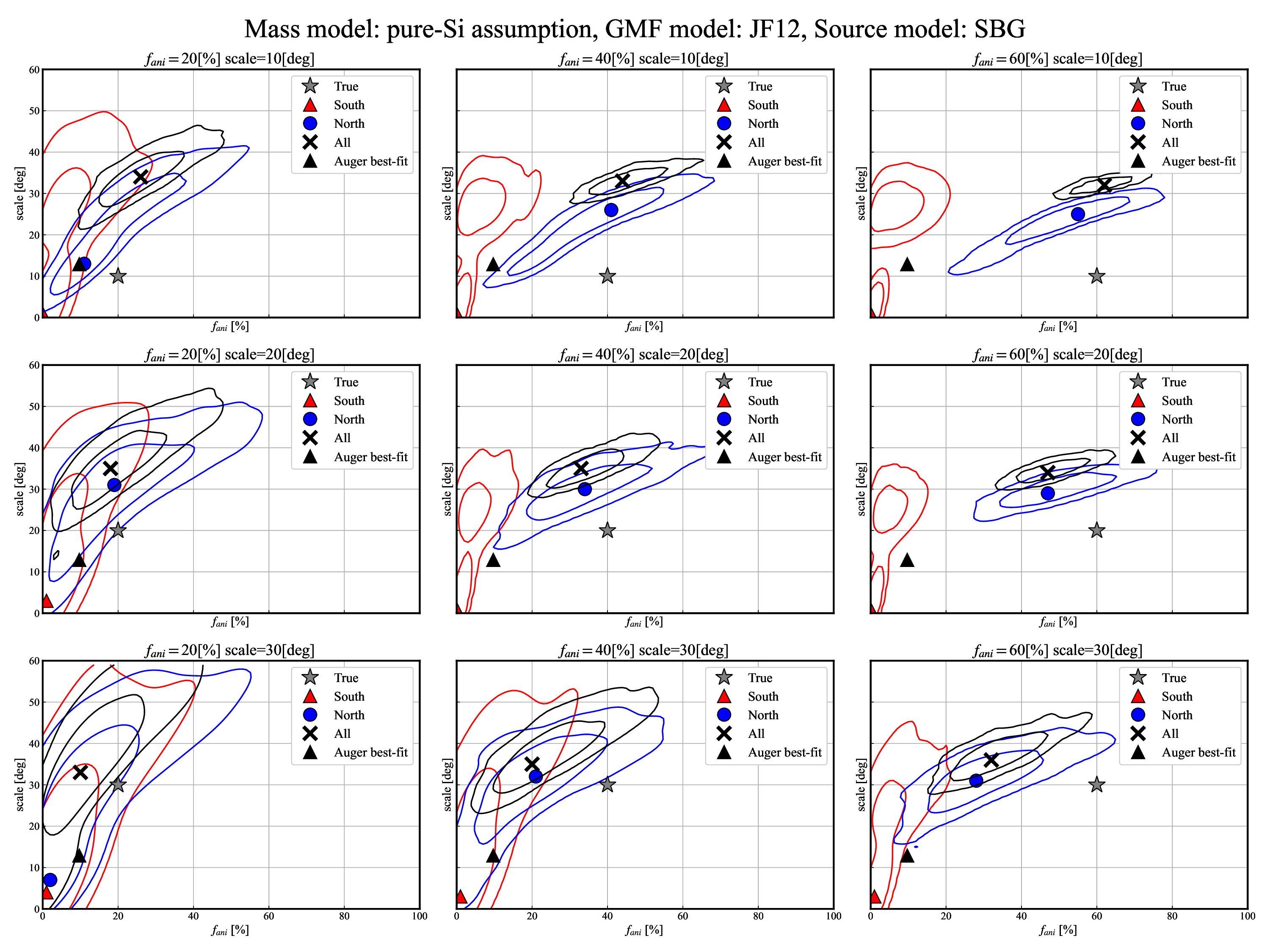}
\caption{
Same as Figure \ref{fig:difpureP} but for the pure-Si case. 
}
\label{fig:difpureSi}
\end{figure}
\begin{figure}
\centering
\includegraphics[width=1\linewidth]{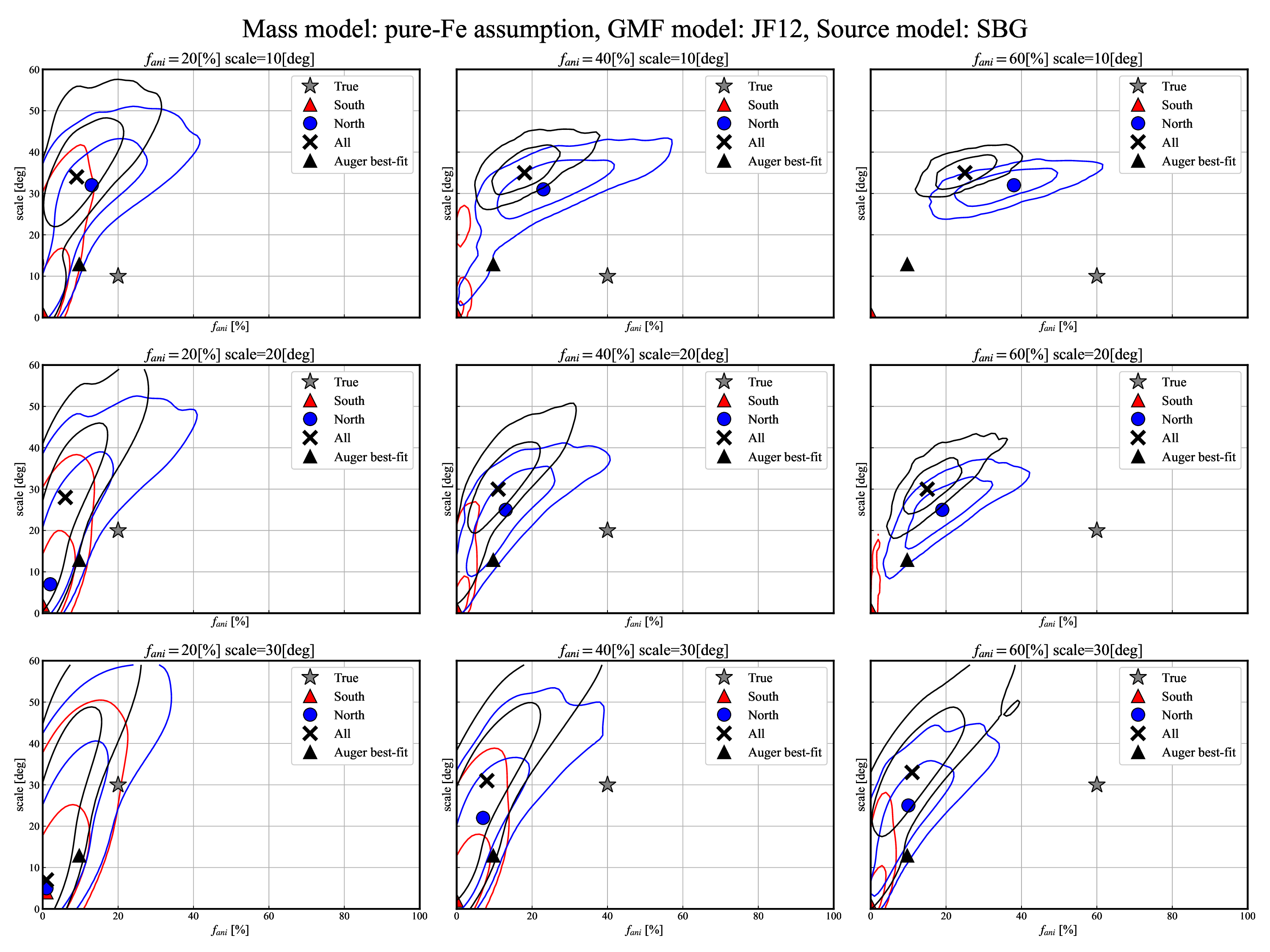}
\caption{
Same as Figure \ref{fig:difpureP} but for the pure-Fe case. 
}
\label{fig:difpureFe}
\end{figure}
The results of the likelihood analysis for single-mass models are shown in Figures \ref{fig:difpureP}-\ref{fig:difpureFe}.
In the pure-proton and pure-He cases, the most frequent best-fit values $(\tilde{f}_{\rm ani},\tilde{\theta})$ fall near the true values $(f_{\rm ani}^{\rm true},\theta^{\rm true})$, which means that the GMF bias is small. 
However, when the separation angular scale $\theta^{\rm true}$ is larger, the dispersion of the estimated parameters, especially for $f_{\rm ani}$ becomes larger.
This dispersion gives an intrinsic statistical uncertainty in the $f_{\rm ani}$ estimation.
In the pure-C case (Figure \ref{fig:difpureC}), a discrepancy in the distributions between the north-sky and south-sky datasets can be seen. 
This tendency becomes larger for the heavier single-mass cases.
In the pure-Si and pure-Fe cases, the most frequent value of $\tilde{f}_{\rm ani}$ in the south-sky datasets becomes $0\,\%$ with any values of the true parameters (Figures \ref{fig:difpureSi} and \ref{fig:difpureFe}). Although both are separated from the true parameters, the distributions of north-sky datasets are closer to those of the all-sky datasets in any case. 
The single and dominant source contribution of M82 and smaller deflection by the GMF in the northern sky can explain this tendency. 

\section{Analysis with PT11 model}\label{app:A1}
\begin{figure}
	\centering
	\includegraphics[width=1\linewidth]{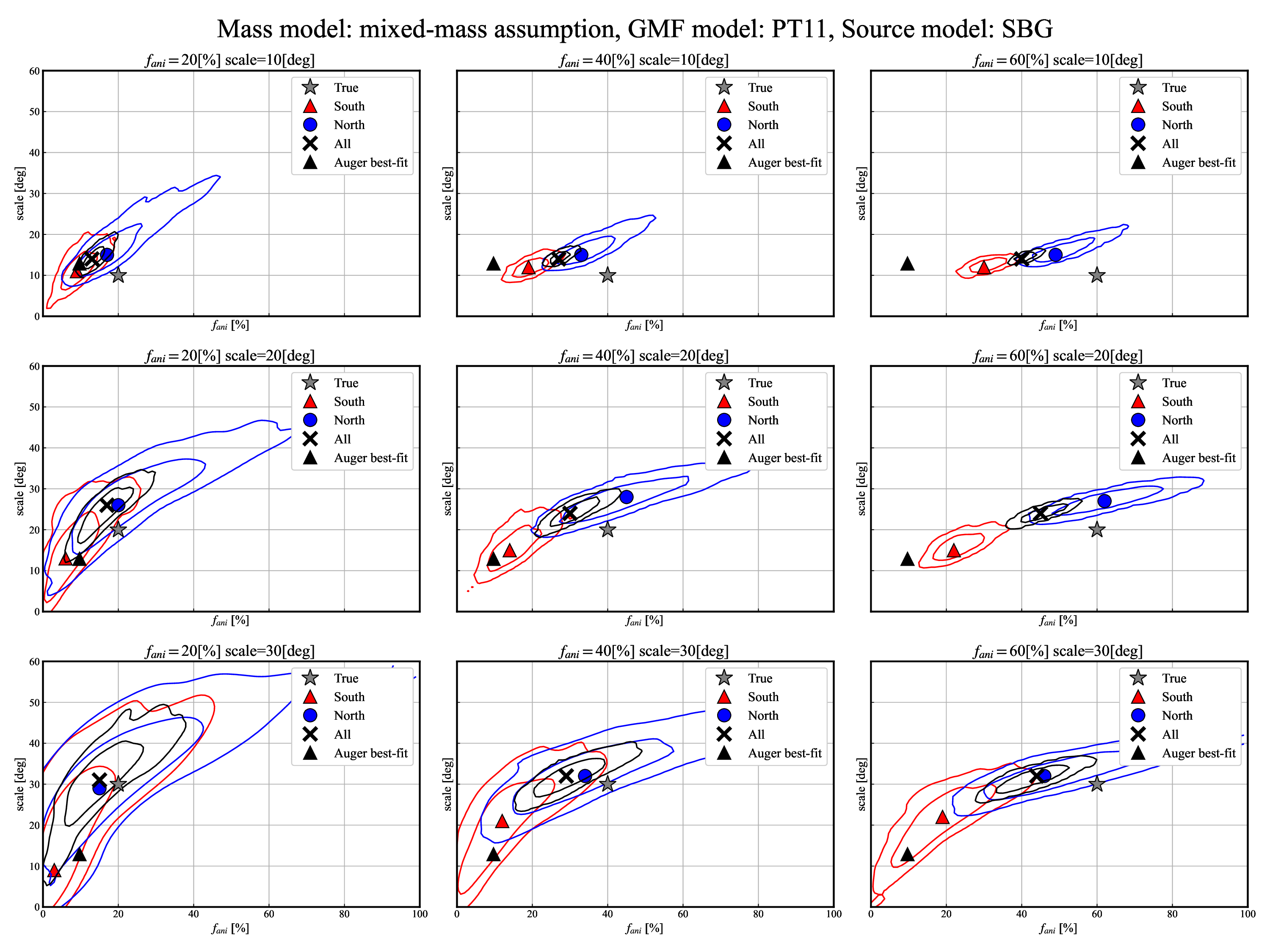}
	\caption{
Same as Figure \ref{fig:difMMCv3} but for the mock event datasets generated with the PT11 model and the mixed-mass assumption. 
	}
	\label{fig:difPT11MMCv3}
\end{figure}
For independent comparison with the JF12 model, we also refer to the  Pshirkov $\&$ Tinyakov 2011 model (PT11) in this study \citep{Pshirkov2011}. 
The mock event datasets are generated in the same manner as in Section \ref{sec:genmmc}, except for the GMF model. 
Figure \ref{fig:difPT11MMCv3} shows the distributions of best-fit parameters in the same manner as Figure \ref{fig:difMMCv3}.
Although there is a quantitative difference, both results show the same tendency due to the GMF bias. 

\bibliography{ms}{}
\bibliographystyle{aasjournal}

\end{document}